\documentclass[twocolumn,prc,aps,showpacs,amsmath,amssymb]{revtex4}
\usepackage{graphicx}

\def\eq#1{{Eq.~(\ref{#1})}}
\def\fig#1{{Fig.~\ref{#1}}}
 \newcommand\la{\langle}
 \newcommand\ra{\rangle}
 \newcommand\beq{\begin{equation}}
 
 \newcommand\eeq{\end{equation}}                                               
 \newcommand\beqn{\begin{eqnarray}}
 \newcommand\eeqn{\end{eqnarray}}
\newcommand{\as}{\alpha_S}
\newcommand{\Lb}{\left(}
\newcommand{\Rb}{\right)}

\def\fm{\,\mbox{fm}}
\def\GeV{\,\mbox{GeV}}

\def\Pom{{\bf I\!P}}

\def\lsim{\mathrel{\rlap{\lower4pt\hbox{\hskip1pt$\sim$}}
    \raise1pt\hbox{$<$}}}         
\def\gsim{\mathrel{\rlap{\lower4pt\hbox{\hskip1pt$\sim$}}
    \raise1pt\hbox{$>$}}}         

\begin{document} 
 
\title{Unitarity bound for gluon shadowing}

\author{B.Z.~Kopeliovich$^{1,2}$}
\author{E.~Levin$^3$}
\author{I.K.~Potashnikova$^1$}
\author{Ivan~Schmidt$^1$}
 
 \affiliation{$^1$Departamento de F\'\i sica
y Centro de Estudios
Subat\'omicos,\\ Universidad T\'ecnica
Federico Santa Mar\'\i a, Casilla 110-V, Valpara\'\i so, Chile\\
{$^2$Joint Institute for Nuclear Research, Dubna, Russia}\\
 {$^3$Department of Particle Physics, School of Physics and Astronomy,
Raymond and Beverly Sackler Faculty of Exact Science, Tel Aviv University, Tel Aviv, 69978, Israel }}
 
\date{\today}
\begin{abstract}
Although at small Bjorken $x$ gluons originated from different nucleons in a nucleus overlap in the longitudinal direction, most of them are still well separated in the transverse plane, therefore cannot fuse. For this reason the gluon density in nuclei cannot drop at small $x$ below a certain bottom bound, which we evaluated in a model independent manner assuming the maximal strength of gluon fusion. We also calculated gluon shadowing in the saturated regime using on the Balitsky-Kovchegov equation, and found the nuclear ratio to be well above the unitarity bound.
The recently updated analysis of parton distributions in nuclei \cite{eps08} including
RHIC data on high-$p_T$ hadron production at forward rapidities, led to astonishingly strong gluon shadowing, which is far beyond  the unitarity bound. This indicates a misconception in the interpretation of the nuclear suppression observed at HRIC.

\smallskip
 
\end{abstract}
\maketitle
 \section{Introduction}
Gluon shadowing, or suppression of the gluon density at small Bjorken $x$ in nuclei, keeps challenging the high-energy-physics community. Lacking direct information from data, attempts have been made to extract gluon density from data on deep-inelastic scattering (DIS) on nuclei. Parton distribution functions (PDF) for different species are related to each other via the DGLAP evolution, and the $Q^2$ dependence of the quark distribution in nuclei should contain information about the gluon density. 
We think that the best analysis of this kind done so far was performed in the next-to-leading order (NLO) by De Florian and Sassot \cite{florian}, who found good sensitivity of data to the gluon distribution and concluded with a rather weak effect of gluon shadowing (as was predicted in \cite{kst2,kpps}).

However, a word of caution is in order. This kind of analysis cannot be considered 
as sefconsistent from the theoretical point of view. Indeed, an NLO analysis has to include an equation for gluon shodowing with $\alpha_s^2$ accuracy. However,  such an equation has not been found so far, therefore NLO analyses \cite{florian,kumano2} are not reliable. 

A self-consistent theoretical scheme has been formulated so far only for a leading order (LO) analysis. Unfortunately,
previous attempts to obtain the gluon distribution in nuclei employing a LO procedure  \cite{eks98,kumano1} failed since the  data from the New Muon Collaboration (NMC) experiment involved in the analyses spans only down to $x\approx 10^{-2}$ and do not
have sufficient accuracy to be sensitive to the gluon distribution (the frequently used gluon shadowing from \cite{eks98} is just an educated guess).

\begin{figure}[htb]
 \includegraphics[width=7cm]{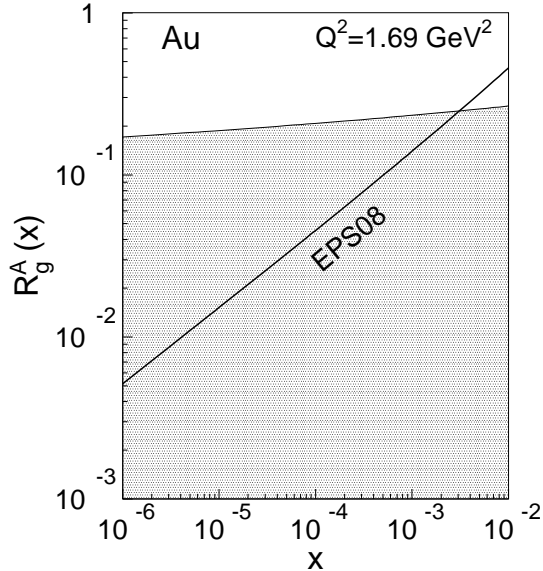}
 \caption{\label{r-x-au} Gold to proton ratio of gluon densities at $Q^2=1.69\GeV^2$ as function of Bjorken $x$. The thin curve shows the bottom unitarity bound for the gluon ratio $R^A_g(x)$, and the shaded area covers the values forbidden by this bound. 
  The thick solid curve is calculated with the EPS08 code \cite{eps08}.}
 \end{figure}

In order to reach the smallest values of $x$ in nuclei one could rely on data for hard processes in hadron-nucleus collisions at forward rapidities. Indeed, the BRAHMS experiment  \cite{brahms} at RHIC performed measurements of high-$p_T$ pion production in $d-Au$ collisions at $\sqrt{s}=200\GeV$ and pseudo-rapidities $\eta=2.2$ and $\eta=3.2$, and the STAR experiment \cite{star} reached $\eta=4$,
in the deuteron fragmentation region.
Although these data allow to access a new range of very small $x$ in the nucleus,
another word of caution is in order: QCD factorization breaks down close to the kinematic limit \cite{knpjs}. This happens because the phase space for partons in the target does not rise, but shrinks (both for proton or nuclear targets), and because of initial/final state interactions in the nucleus. Indeed, inclusion of these data 
in the new global analysis \cite{eps08} resulted in an astonishingly small amount of gluons in nuclei.
Fig.~\ref{r-x-au} shows the result for the gold to proton ratio of gluon densities $R_g^A(x,Q^2)$ as function of $x$ at $Q^2=1.69\GeV^2$. 
 According to \cite{eps08} gluons in the nucleus are suppressed by orders of magnitude at small $x$. This implies that
the effective number of nucleons,
\beq
A_{eff}=A\,R^A_g(x,Q^2),
\label{10}
\eeq
is extremely small. At $x=10^{-6}$ the new nuclear PDF \cite{eps08} predicts $A_{eff}=1$, i.e. the whole gold nucleus is represented by a single nucleon (!). Certainly, this cannot be true, unless all nuclear gluons are strongly correlated in impact parameter plane throughout the whole nucleus. Such a strong gluon shadowing also contradicts the bottom unitarity bound derived below, which is plotted in Fig.~\ref{r-x-au} by the thin curve. The nuclear ratio resulted from the EPS08 analysis is fully in the shaded area forbidden by the bound.

This controversy indicates a misconception in the interpretation of the suppression effect observed by the BRAHMS experiment. This also explains the failure of the analysis \cite{eps08} to describe data \cite{brahms} at $p_T^2<4\GeV^2$ and similar data \cite{star} at larger rapidity $\eta=4$.

Actually, the effect of enhanced nuclear suppression at forward rapidities observed by the BRAHMS experiment is not unique. It is known to happen in every reaction measured so far, even at low energies, where no coherent effects are possible (see examples in \cite{knpjs}). Although at forward rapidities one accesses the smallest $x$ in the nuclear target, one simultaneously gets into the region of large $x$ of the beam hadron, where energy conservation becomes an issue. The projectile hadron, or its remnants, propagating through the nucleus dissipate energy for multiple interactions in the nucleus. This makes it more difficult to give the main fraction of the initial energy to one particle detected with large Feynman $x_F$ or/and transverse $x_T$. Consequently, multiple interactions enhance nuclear suppression at forward rapidities. 
Calculations performed in \cite{knpjs} based on the AGK cutting rules well describe the BRAHMS data \cite{brahms} and correctly predict \cite{jan} suppression at larger rapidity \cite{star}, as well as in other reactions measured at large 
$x_F$. This nuclear suppression scales (approximately) in Feynman $x_F$, rather than in $x_2$ as should have been expected for nuclear shadowing. 
Notice that the effective energy loss is a nonperturbative effect, which can be evaluated only within phenomenological models.

The effect of energy dissipation caused by multiple collisions should not be mixed up with gluon shadowing which is also related to multiple interactions, but is responsible only for a 
part of the observed nuclear effects.  
Attributing the whole suppression to gluon coherence effects (shadowing or color glass) leads to grossly exaggerated gluon shadowing \cite{eps08}  which, as we will show below,  breaks the unitarity bound. Therefore,   the derivation of the minimal value for $R^A_g$  may  give a useful guide for the separation of the two effects, gluon shadowing and energy loss, in the analysis of experimental data. In order to discriminate the two mechanisms, one can also move to lower energies where coherence phenomena disappear, while the effects of energy conservation, which scale in $x_F$, remain essentially unchanged.

\section{Gluon saturation}

Shadowing, both the term and the phenomenon, which came from optics, got a new interpretation within the parton model \cite{kancheli}. 
Although the bound nucleons are well separated in the nuclear rest frame, they start "talking" to each other being boosted to the infinite momentum frame. At first glance both the inter-nucleon 
spacing and the nucleons themselves are subject to Lorentz contraction with the same $\gamma$-factor, $\gamma_N=p_N^+/m_N$. So they should remain separated. However, a boosted nucleon develops quantum fluctuations, sea partons, which are an analogue to Weizs\"acker-Williams photons.
A parton carrying a fraction $x$ of the proton light-cone momentum contracts less, since has a smaller $\gamma$-factor, $\gamma(x)=x\gamma_N$. Thus, the parton clouds originated from different nucleons should overlap at small $x<<1$ and may fuse leading to a reduction of the parton density. The corresponding  nonlinear fusion term was introduced in the Gribov-Levin-Ryskin \cite{grl} and Mueller-Qiu  \cite{mq} (GLR-MQ) equation,
\beqn
\frac{\partial^2xG_A(x,Q^2,b)}
{\partial\ln(1/x)\partial\ln(Q^2)} &=&
\frac{\alpha_sN_c}{\pi}\,
xG_A(x,Q^2,b)
\nonumber\\ &-&
\frac{4\pi\alpha_s^2N_c}{3C_FQ^2}\,
\left[xG_A(x,Q^2,b)\right]^2.
\label{100}
\eeqn
Here 
\beq
G_A(x,Q^2,b)\equiv\frac{d}{d^2b}G_A(x,Q^2),
\label{200}
\eeq
is the impact parameter density of the nuclear gluon distribution function. In absence of the nonlinear term in (\ref{100}) the unshadowed nuclear gluon density can be related to the nucleon one as, 
\beq
G^0_A(x,Q^2,b)=T_A(b)\,G_N(x,Q^2),
\label{300}
\eeq
where $T_A(b)=\int_{-\infty}^{\infty}dz\,\rho_A(b,z)$ is the nuclear profile function, the integral of the nuclear density along the hadron trajectory.
Thus, the gluon density in heavy nuclei would rise $\propto A^{1/3}$, but the nonlinear term rises faster, $\propto A^{2/3}$, and slows down the $A$-dependence. Notice that the GLR-MQ equation with a quadratic nonlinear term describes only the onset of saturation. While $G_N(x,Q^2)$ increases, the higher order terms become important.
Eventually, at large $A$ the gluon density is expected to reach a saturated value $G_A^s(x,Q^2, b)$, which is independent of $b$ (except at the nuclear periphery) .  Thus, in the saturation limit the nuclear ratio of gluon densities
depends on $A$ as,
\beq
R_g^s(x,Q^2)=\frac{\int d^2b\,G_A^s(x,Q^2,b)}
{A\,G_N(x,Q^2)}\propto A^{-1/3}.
\label{302}
\eeq

The same phenomenon of saturation can be understood in the rest frame of the nucleus, which gives a more familiar interpretation having direct ties to the usual interpretation of shadowing in optics. In this frame gluon shadowing can be treated as Glauber shadowing for a glue-glue dipole \cite{al}, or as a contribution of higher Fock components 
of the projectile containing gluons \cite{kst2}. In this frame saturation corresponds to the black disk limit for the dipole-nucleus amplitude.

\section{Model independent  unitarity bound}
 
 There must be a bottom bound for the saturated density, i.e. for the strength of gluon shadowing. Indeed, how many gluons are left in a heavy nucleus at small $x$ after many fusions? Can it be as little as in one nucleon? As far as the nucleon is much smaller than the nucleus, this is obviously impossible. Two nucleons separated in impact parameter plane by a long distance $\sim R_A$ do not "talk to each other" even at small $x$ 
 \cite{footnote}. There should be some minimal amount of gluons at small $x$, corresponding to a maximal fusion rate at a given impact parameter.
Assume that we have a one-dimensional row of nucleons.  Intuitively, even at very low $x\ll1/m_NR_A$, where all gluons originated from different nucleons, overlap and fuse, the resulting gluon density cannot be smaller than in a single nucleon. Indeed, upon approaching the gluon density in a single nucleon, the further evolution in $x$ described by  Eq.~\ref{100} proceeds exactly like in a nucleon.
So it is clear that the minimal amount of gluons in the nucleus should be proportional to its area, i.e. to $A^{2/3}$, but the coefficient is important.

Gluons originating from different nucleons and overlapping in longitudinal direction can fuse if their transverse separation is within the gluon interaction radius $r_g(x)$. The latter can be interpreted as the radius of the triple-Pomeron vertex.  If the radius of gluon distribution in the nucleon is $r_N(x)$,
the gluon clouds of two nucleons overlap, if the distance between the nucleon centers does not exceed $r_{NN}=\sqrt{r_N^2(x)+r_g^2(x)}$.
The mean number of nucleons whose gluons are able to fuse at impact parameter $b$ is,
\beq
\la n(b)\ra= \pi r_{NN}^2\,T_A(b).
\label{310}
\eeq
Assuming uncorrelated  Poisson distribution, the probability to find $n$ nucleons  in the tube is  $W_n=\exp(-\la n\ra)\,\la n\ra^n/n!$. Maximal gluon shadowing corresponds to full fusion of gluons belonging to different nucleons in the tube, i.e. the collective gluon density of $n\geq1$ overlapping nucleons equals to a single nucleon gluon density, $nG_N(x,Q^2)\Rightarrow G_N(x,Q^2)$. 
The probability to have one or more nucleons in the tube, i.e. to have a saturated gluon density,  is $1-\exp\bigl(-\la n\ra\bigr)$. Correspondingly,
the shadowed gluon density corrected for finite $A$ has the form,
\beq
\tilde G_N(x,Q^2)=G_N(x,Q^2)\,
\left\{1-\left[1-\frac{\la n(b)\ra}{A}\right]^A\right\},
\label{320}
\eeq
and the nuclear gluon density  per unit of transverse area is
\beq
G_A(x,Q^2, b)=\frac{\tilde G(x,Q^2)}{\pi r_{NN}^2}.
\label{330}
\eeq

 Correspondingly, the bottom bound for the nucleus-to-nucleon ratio of gluon densities reads,
\beq
R_g^A(x,Q^2)\equiv \frac{\int d^2b\, G_A(x,Q^2,b)}
{A\,G_N(x,Q^2)}\geq
\frac{\sigma_A^{eff}}{A\,\sigma_N^{eff}},
\label{340}
\eeq
where $\sigma_N^{eff}=\pi r_{NN}^2$; $\sigma_A^{eff}=\int d^2b\bigl\{1-\bigl[1-{1\over A}\sigma_N^{eff}T_A(b)\bigr]^A\bigr\}$. Here we maximized the denominator, assuming fully saturated gluon density in a nucleon.
For heavy nuclei this bound Eq.~(\ref{340}) can be approximated as $R_g^A(x,Q^2)\geq (R_A^2/A\,r_{NN}^2)\propto A^{-1/3}$.

Now we should settle the values of $r_g$ and $r_N$ which are functions of $x$ and $Q^2$. Here we should rely on available experimental data. Gluons originated from two nucleons with different impact parameters overlap, if the mean distance squared, $r_{N}^2(x,Q^2)$,
equals to the doubled mean radius squared of the gluon distribution in each proton. 
The latter was measured at HERA in reaction of virtual electroproduction of vector mesons,
\beqn
{1\over2}\, r_N^2(x,Q^2)&=&2B_{el}(x,Q^2)
\nonumber\\&=&
2B_0(Q^2)+4\alpha_\Pom^\prime(Q^2)\ln(1/x).
\label{350}
\eeqn
The $x$-independent term $B_0\approx 4\GeV^{-2}$ is related to the proton formfactor; the effective slope of the Pomeron trajectory $\alpha^\prime_\Pom(Q^2)$ varies with $Q^2$. It is known to be as large as $0.25\GeV^{-2}$ in soft hadronic processes, and is essentially a result of unitarity saturation at small impact parameters \cite{kp,kpps}. Data on exclusive electroproduction of vector mesons show that at high $Q^2$ the effective slope is much smaller, $ \alpha^\prime_\Pom(Q^2)\approx0.1\GeV^{-2}$ \cite{slope1,slope2} (see more references in \cite{kpps}). To be on the safe side (regarding model independence of the bottom bound) we use the maximal value $\alpha^\prime_\Pom=0.25\GeV^{-2}$.

The radius of the triple-Pomeron vertex, $r_g(x)$, is well measured in soft diffractive dissociation
of protons to large invariant masses $pp\to pX$ available with high statistics. The triple-Regge analysis of data \cite{kklp} shows that this radius is compatible with zero, in any case is much smaller than $r_N$. This observation gets a natural explanation\cite{kpps} if the mean transverse momentum of gluons in the proton is large and controlled by a saturation scale $Q_s(x)$ which increases as a function of energy \cite{grl,mq,MV}.
Therefore,  the radius $r_g(x)$ can only shrink, so we can safely neglect it compared to the nucleon radius $r_N$.

With these values of the parameters the bottom unitarity bound, Eq.~(\ref{340}), for gluon suppression in gold at $Q^2=1.69\GeV^2$ is depicted in Fig.~\ref{r-x-au}. The nuclear ratio $R_g^A$ calculated with EPS08 is also plotted in this figure and is fully in the shaded area below the unitarity bound.

\begin{figure}[htb]
 \includegraphics[width=7cm]{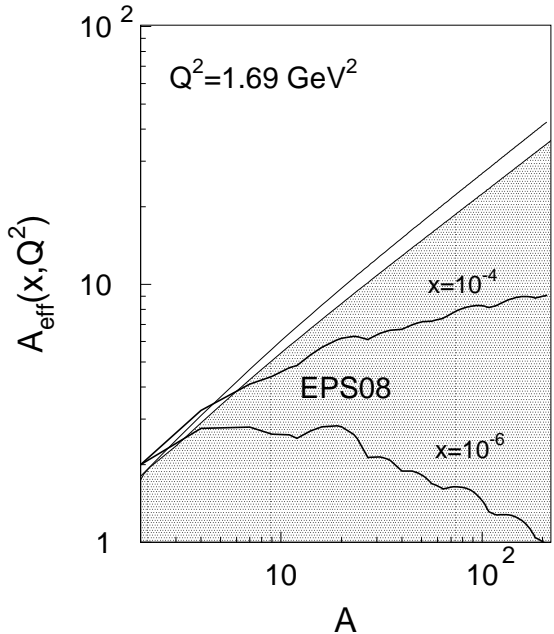}
 \caption{\label{a-dep} Unitarity bounds for the effective number of nucleons, $A_{eff}=A\,R_g^A$, as function of $A$ calculated with Eq.~(\ref{340}) at $Q^2=1.69\GeV^2$ for $x=10^{-4}$ (top thin line) and $x=10^{-6}$ (bottom thin line). Solid thick curves are calculated  with the EPS08 code \cite{eps08}.}
 \end{figure}

One can also trace the $A$-dependence of gluon shadowing. In Fig.~\ref{a-dep} we plotted the effective number of nucleons, $A_{eff}=A\,R_g^A(x,Q^2)$, suggested by the EPS08 analysis at $Q^2=1.69\GeV^2$ and $x=10^{-4},\ 10^{-6}$.
$A_{eff}$ is quite small, especially at $x=10^{-6}$ where for gold, as we already mentioned, $A_{eff}=1$. The $A$ dependence also looks odd: $A_{eff}$ never falls with $A$, it can only rise, as $A^{2/3}$ or faster. Apparently, the predicted $A$-dependence at $x=10^{-6}$ is unacceptable. Although it rises with $A$ at $x=10^{-4}$, but too slow, breaking the unitarity restriction. 

We also calculated the $A$-dependent unitarity bounds, Eq.~(\ref{340}),  and plotted the 
results  in Fig.~\ref{a-dep} at $x=10^{-4}$ and $x=10^{-6}$ by thin solid lines, top and bottom respectively. Again, the results of EPS08 shown by thick solid curves are quite below the bounds.

\section{More stringent constraints}

Trying to be model independent we apparently exaggerated some estimates. We assumed 
in (\ref{320}), (\ref{340}) that gluons in different nucleons completely overlap in longitudinal direction, i.e. that the coherence length is much longer than the nuclear size. This is questionable even at very small $x$ (due to high-mass gluonic fluctuations), and is certainly not the case, say, at $x\sim 10^{-2}$ where the coherence length for gluons is shorter than the nuclear radius \cite{krt}. In this range of $x$ the bottom bound should be considerably lifted up. Actually, the onset of gluon shadowing starts only at $x\lsim 10^{-2}$ \cite{kst2}.  We exaggerated the strength of nuclear shadowing for the unitarity bound in order to keep it model independent.

We also maximized the strength of gluon fusion, assuming that the appearance of an extra gluon results in an immediate fusion, so that the gluon density remains unchanged. This corresponds to a black disk limit for $pp$ elastic partial amplitude, or full saturation, while data show \cite{k3p} that this regime  is relevant only to central ($b\approx0$) $pp$ collisions.

Then, we maximized the glue-glue correlation radius, which we evaluated at $r_N\approx 1\fm$, while calculations on the lattice lead to a much smaller value of about $0.2-0.3\fm$ \cite{lattice}. There are also numerous evidences in data \cite{kp,kpps} indicating that gluons are located within small spots of radius $r_0\approx 0.3\fm$ around the valence quarks. Although, following the valence quarks, they are distributed in impact parameters with large radius Eq.~(\ref{350}), the probability to overlap in impact parameters is significantly reduced. 
Indeed, for a given gluonic spot the mean number of others to overlap with is,
\beq
\la n_g\ra=\pi r_0^2\,3\la T_A\ra,
\label{400}
\eeq
where the mean nuclear thickness, $\la T_A\ra={1\over A}\int d^2b\,T_A^2(b)$ is about $1.5\fm^{-2}$ for heavy nuclei. Thus, transverse overlap of gluonic spots is quite small even for heavy nuclei, $\la n_g\ra\approx1.3$. 
To improve the unitarity bound one can rely on Eq.~(\ref{340}), but with a different
value of the effective cross section, $\sigma_{eff}^N\Rightarrow\sigma_{eff}^q=3\pi r_0^2$.
The new bound for gold at $Q^2=0.169\GeV^2$, $R^A_g>0.6$, is about three times higher than the model independent one, depicted in Fig.~\ref{r-x-au}.\\

\section{Gluon shadowing from the Balitsky-Kovchegov (BK) equation.}

It is interesting to compare the unitarity bound for gluon shadowing with expectations coming from contemporary models of gluon saturation in nuclei.
Although the GLR-MQ equation (\ref{100}) describes  qualitatively the phenomenon of gluon saturation and allows to take explicitly into account the nuclear target, the correct equation that governs the shadowing effects  (BK equation \cite{BK} ) has a more complicated form, namely,
\begin{eqnarray}
&&
\frac{\partial N\Lb r,Y;b \Rb}{\partial\,Y}\,
=\,\,\frac{C_F\,\as}{\pi^2}\,\,\int\,\frac{d^2 r'\,r^2}{(\vec{r}\,-
\,\vec{r}\:')^2\,r'^2}\, \label{BK}\\
&&
\left[ \,2  N\Lb r',Y;\vec{b} \, -\ 
\frac{1}{2}\,(\vec{r} - \vec{r}\:')\Rb\,\,- \,\, N\Lb r,Y;\vec{b}\Rb\,\,\right.\notag\\
&&\left. - \,\,
 N\Lb r',Y;\vec{b} - \frac{1}{2}\,(\vec{r} - \vec{r}\:') \Rb\, N\Lb
 \vec{r} - 
\vec{r}\:',Y;b - \frac{1}{2} \vec{r}\:'\Rb  '
\right] \notag
\end{eqnarray}
where $Y =\ln(1/x)$. This equation  has a form different from  the GLR-MQ, since the nuclear target enters only in the initial condition
at some rapidity $Y = Y_0$. This happens because the equation  describes a simple process: a small dipole of size $r$ decays into two dipoles of sizes $r'$ and $|\vec{r} - \vec{r}^{\,\prime}|$ with probability
$K(r,r')\,=\,\frac{C_F\,\as}{\pi^2}\,\,\frac{r^2}{(\vec{r}\,-
\,\vec{r}\:')^2\,r'^2}$, which does not depend on the target.  These two dipoles can interact either independently 
( the linear term of the equation), or
simultaneously  ( the non-linear term of the equation).  In the case of the nuclear target one can simplify \eq{BK} taking $b \approx R_A \,\gg\, {r'} \,\mbox{and} \,|\vec{r} - \vec{r}^{\,\prime}|$. So $b$ becomes a parameter in the equation, and the $b$-dependence is entirely determined by the initial condition. In other words, there is no correlations between partons with different $b$. It should be noticed that an analysis of HERA data based on the BK equation was performed in the entire kinematic range of $Q^2$ from $Q^2 = 0.1\,GeV^2$ to $Q^2 =100 \,GeV^2$ with excellent $\chi^2/d.o.f <  1 $ \cite{HERAAN}.

 The key property of the BK equation (as well as of the GLR-MQ one) is the appearence of the new dimensional scale $Q_s(x;b)$ \cite{grl,MV,LT}.
 For further discussion it is worth mentioning that the saturation momentum for nuclei equals to \label{KLM,MUQS,LT}
\beq \label{QSA}
Q_s\Lb A; x; b\Rb \,\,=\,\,\int d^2 b'\, Q_s\Lb N; x; b'\Rb\,T_A(b)
\eeq
It turns out that inside the saturation domain and even outside, the solution of the BK equation $N(r,Y;b)$ obeys geometrical scaling \cite{GS}, i.e. depends only on one variable,
$rQ_s(x,b)$.
In Ref. \cite{LT}  it was found that in the saturation domain it has the form,
\beqn \label{LT}
&&N(r,Y;b) \,\,=\,\,1 - C \,\exp\left[-  \frac{1 - \gamma_{cr}}{2 \,\chi(\gamma_{cr})}\,\,z^2\right], \eeqn
where $z=\ln \Lb r^2 \,Q^2_s(x;b)/4\Rb$,
and $\chi(\gamma)$ is the BFKL kernel (function $K(r,r')$ in $\gamma$, i.e. anomalous dimension) representation; $\gamma_{cr}$ is the critical value of anomalous dimension corresponding to the saturation momentum (see details e.g. in Ref. \cite{grl}).
\eq{LT} shows how  approaching the unitarity boundary depends on the target, namely, through its saturation momentum.
Unfortunately , the value of the coefficient $C$ in \eq{LT} can be taken only from the numerical solution of the BK equation, and it was found in   Ref. \cite{LT} to be equal to $C=1/e$.
Further, the ratio $R^A_g$ can be written in the form
\beq \label{BK1}
R^A_{g}\Lb x,Q\Rb\,\,=\,\,\frac{\int d^2 r |\Psi_g(Q;r)|^2\,N_A\Lb r,x;b\Rb}{A\,\int d^2 r |\Psi_g(Q;r)|^2\,N_p\Lb r ;x;b\Rb}
\eeq
where $|\Psi_g(Q;r)|^2$ is the probabity to find a $g g$ dipole in the probe. This wave function has been discussed in \cite{WF,kst2}. It turnes out that $|\Psi_g(Q;r)|^2 \,\propto\,\frac{1}{Q^2\,r^4}$ for $r > 1/Q$. For $N$ we use \eq{LT} with the saturation momentum for a nucleus given by \eq{QSA}, and for a nucleon we take
\beq \label{QSN}
Q_s\Lb N; x;b\Rb \,\,=\,\,Q_s(x) \,S(b)
\eeq
with 
$$ S(b) \,=\,\, \Lb \frac{2 \sqrt{2}\,b}{R}\Rb \,\,K_1\Lb \frac{2 \sqrt{2}\,b}{R}\Rb$$
One can see that $S(0)$ =1 and with $R=\sqrt{\la r_{ch}^2\ra}=0.89\,\, fm$ it gives correct electromagnetic radius of the proton.

For $N(r,Y;b)$ in \eq{BK1} we take \eq{LT} with $r^2 Q_s(x,b)/4 >1$. Therefore, in \eq{BK1}  we   should integrate over  $r > r_{min}$ where
\beqn \label{BK3}
r^N_{min}\,\,&=&\,\,\mbox{maximum} \left\{ \frac{2}{Q} \,\,\mbox{or} \,\,\frac{2}{Q_s\Lb N; x;b\Rb}\right\}\,;\,\,\,\,\,\notag\\
r^A_{min},\,&=&\,\,\mbox{maximum} \left\{\frac{2}{Q} \,\,\mbox{or} \,\,\frac{2}{Q_s\Lb A ; x;b\Rb}\right\}\,;\,\,\,\,\,
\eeqn

\begin{figure}[t]
 \includegraphics[width=7cm]{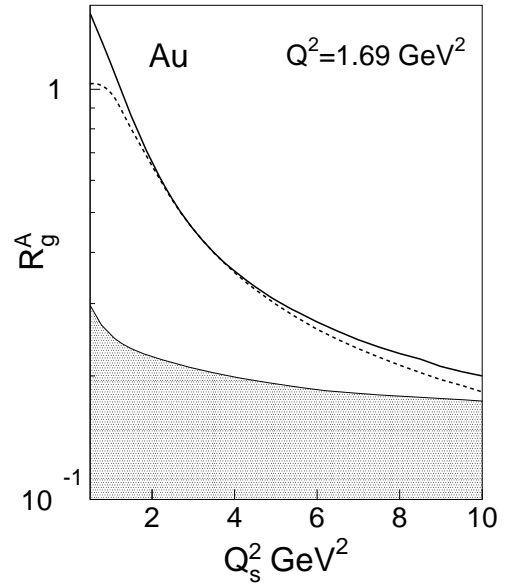}
 \caption{Gold to proton ratio of gluon densities $R^A_g$ at $Q^2=1.69\GeV^2$ as function of  saturation momentum squared $Q^2_s(x)$ (see Eqs.~\ref{QSA} and \ref{QSN}) for two models for the dipole amplitude $N(r,x;b)$: solid curve for 
the solution of the BK equation in the saturation domain (see \eq{LT} ); and dashed curve for the phenomenological GBW model (see \eq{GW}).  \label{bkra}}
 \end{figure}

\begin{figure}[t]
\includegraphics[width=7cm]{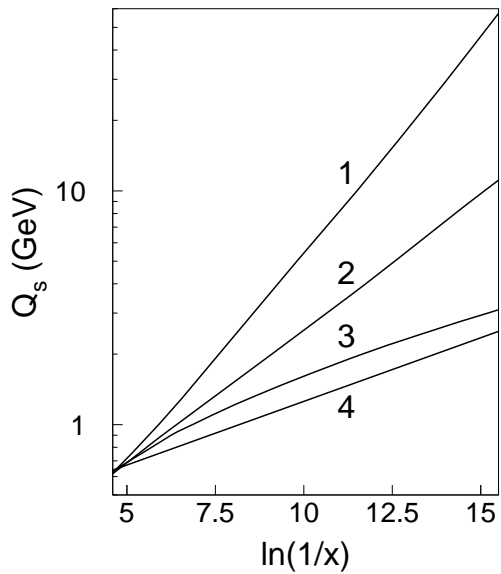}
 \caption{ Saturation momentum $Q_s(x)$ for the proton versus $\ln(1/x)$ . Curves 1 and 2 describe  $Q_s$ with LO and NLO BFKL kernels (see Ref. \cite{DGQS}) respectively. Curve 3 presents the saturation momentum that stems from the modified BK equation of \cite{GLMMOD}. Curve 4 is the GBW phenomenological 
saturation momentum. \label{qs} }
 \end{figure}

 The predicted ratio is shown in \fig{bkra} as a function of $Q_s \equiv  Q_s(x) $ in \eq{QSA} and \eq{QSN}, while in \fig{qs} we show the prediction for the saturation momentum as function of $\ln (1/x)$ as it was calculated in \cite{GLMMOD}). 
 
  Notice that our calculations are valid only for $r^2>1/Q_s^2$, a condition which is broken at small $Q_s$. This is the reason for the odd effect of antishadowing which one can see in Fig.~\ref{bkra} at small $Q_s$.

We also plot in \fig{bkra}  the prediction for $R^A_g$ from the model for the dipole amplitude $N(r,x;b)$, inspired by the Golec-Biernat and Wusthoff (GBW) parametrization, namely,
\beqn \label{GW}
&&N(r,Y;b)\,\,=\,\,1\,\,-\,\,\exp\Lb - r^2\,Q^2_s(x;b)/4\Rb \notag
\eeqn
This model describes quite well the DIS data of HERA and comparison with this model can illustrate how close are our theoretical estimates to reality. 

\section{Summary}

\begin{itemize}
\item
Although at small $x$ gluons originated from different nucleons in a nucleus overlap in longitudinal direction, most of them are still well separated in the transverse plane, therefore cannot fuse. For this reason the gluon density in nuclei cannot drop at small $x$ below a certain bottom bound which we evaluate assuming maximal strength of gluon fusion.
We found a model-independent unitarity bound for gluon shadowing. Examples for the nuclear ratio of gold-to-proton and the $A$-dependence of the bound are shown in Figs.~\ref{r-x-au} and \ref{a-dep} respectively.

\item
The recent results of the global fit of nuclear PDFs \cite{eps08} lead to a gluon suppression which is orders of magnitude below the unitarity bound. Moreover,  the amount of effective nucleons predicted by EPS08 has an odd $A$-dependence (rising too slow or even falling with $A$), and is also much below the unitarity limit, as is shown in Fig.~\ref{a-dep}.

The source of the problem is rather obvious. Trying to access minimal Bjorken $x$ for nuclear partons, the EPS08 analysis included RHIC data on high $p_T$ hadron production at forward rapidities. However, these data were analyzed in \cite{eps08} assuming that the mechanism of suppression is related only to a modification of the nuclear PDFs, while at least two different mechanisms contribute to the observed nuclear effects.
  
\item
Since maximal gluon shadowing is expected in the regime of gluon saturation, we performed model calculations based on the Balitsky-Kovchegov equation and found that the results presented in Fig.~\ref{bkra} are well above the unitarity bound.  

\end{itemize}

\begin{acknowledgments}

E.L. wishes to thank the  high energy theory group at the University Federico Santa Maria for hospitality and warm atmosphere which made the work pleasant.

 This work was supported in part by Fondecyt (Chile) grants, numbers 1050519, 1050589, 
7080067 and 7080071, by DFG (Germany)  grant PI182/3-1,
 by BSF grant $\#$ 20004019, by
a grant from Israel Ministry of Science, Culture \& Sport, and
the Foundation for Basic Research of the Russian Federation.

\end{acknowledgments}


\begin{thebibliography}{99}
\bibitem{footnote} One may expect that at unreachably small $x$ hadronic interaction radius will become as large as the nuclear size, and all hadronic cross section, including hadron-nucleus, will approach an universal Froissart behavior. In this case the nuclear ration will reach its bottom limit $R^s_g=1/A$. However, at currently available energies, including LHC, we do not observe even an onset of that regime.
\end{thebibliography}
\end{document}